\documentclass[aps,prd,floatfix,nofootinbib,superscriptaddress,twocolumn,showkeys]{revtex4-2}
\usepackage{amsmath}
\usepackage{amssymb}
\allowdisplaybreaks

\usepackage[thicklines]{cancel}
\usepackage{makeidx}
\usepackage{amsfonts}
\usepackage[usenames,dvipsnames]{pstricks}
\usepackage{subfigure}
\usepackage{epsfig}
\usepackage{pst-grad} 
\usepackage{mathtools}
\usepackage[colorlinks,hyperindex]{hyperref}
\usepackage{array}
\usepackage{tikz}
\usepackage{float}
\usetikzlibrary{decorations.markings,math,shapes.misc,patterns,arrows.meta,calc,patterns,angles,quotes,intersections}
\usepackage{allpurpose}
\newcommand {\nn}{\nonumber}
\usepackage{soul}

\hypersetup
{colorlinks,%
citecolor=black,%
linkcolor=black,%
urlcolor=black,%
}

\begin{document}

\title{Weak deflection angle of charged signal in magnetic fields}

\author{Qianchuan Wang}
\address{School of Physics and Technology, Wuhan University, Wuhan, 430072, China}

\author{Junji Jia}
\email[Corresponding author:~]{junjijia@whu.edu.cn}
\address{Department of Astronomy $\&$ MOE Key Laboratory of Artificial Micro- and Nano-structures, School of Physics and Technology, Wuhan University, Wuhan, 430072, China}

\begin{abstract}
We use the perturbative method to study the influence of the magnetic field on the weak deflection angle of charged signals in magnetized stationary and axisymmetric spacetimes within general electromagnetic potentials. 
The deflection angle is expressed as a series expansion of the inverse of the impact parameter $b$, with coefficients determined by the asymptotic expansions of the metric functions and the electromagnetic four-potential. It is found that in general, the deflection angle can always be separated into two parts, the usual gravitational part as for neutral particles, and the electromagnetic part due to the interaction between the (electro)magnetic field and the signal. The leading order of the gravitational, electrostatic (from nonzero spacetime charge) and magnetic (from nonzero magnetic dipole moment) contributions are $b^{-1},\,b^{-1}$ and $b^{-2}$ respectively. The entire electromagnetic part is enhanced by the large specific charge of elementary particles but suppressed by the reciprocal Lorentz factor. The deflection angle result is then applied to three spacetimes with intrinsic or externally enforced magnetic fields. Effects of the magnetic field on the deflection angle from various parameters, including the spacetime spin, magnetic dipole moment and magnetic parameters,  are analyzed. In all these cases, it is found that in the weak deflection limit, these effects agree with the expectation for a Lorentz force; that is, an attractive (or repulsive) one will enlarge (or decrease) the deflection angle. 
\end{abstract}


\date{\today}
\maketitle

\section{Introduction}

The deflection of light rays played a significant role in the validation of general relativity \cite{Dyson:1920cwa}. The gravitational lensing (GL) based on the signal deflection is currently utilized in astronomical observation for various purposes, such as assessing the mass distribution within galaxy clusters \cite{ME:1991,Jullo:2010dq,Kneib:2011jaf,Umetsu:2020wlf}, constraining cosmological parameters \cite{Refregier:2003ct,Hildebrandt:2016iqg,Lorenz:2017iez} and examining alternative theories that extend beyond general relativity \cite{Wu:1998ju,SDSS:2003bun,Acquaviva:2004fv,Cunha:2018acu,Mukherjee:2019wcg}.

In addition to the investigation of light rays, there has been a growing interest in the study of the deflection and GL of various signal types in recent years, such as the gravitational waves \cite{LIGOScientific:2016aoc,LIGOScientific:2017zic,Mishra:2021xzz}, cosmic microwave background \cite{Lewis:2006fu}, neutrinos \cite{Kamiokande-II:1987idp,Harnois-Deraps:2014sva,IceCube:2018cha} and cosmic rays  \cite{Tinyakov:2004pw,Keivani:2014kua,Farrar:2017lhm}. As these signals have been increasingly documented through cosmological observations, their deflection and GL are emerging as significant instruments for cosmological research. 
Among these signals, the cosmic rays are mainly composed of energetic charged particles  \cite{ParticleDataGroup:2018ovx} and their deflection is of particular interest to us.

On the other hand, it is also known that in the vast space of the cosmos, there exist various types of magnetic fields of vastly different strength  \cite{Vallee:2004osq,Durrer:2013pga}. The charged signals, when flying by these magnetic fields, will unavoidably experience deflection and possibly GL by the magnetic field, on top of the bending effect due to massive gravity. 
In particular, it is known that the Earth is positioned within a vast magnetic ring encircling the core of the Milky Way \cite{Beck:2016,Unger:2023lob}.
Moreover, recently there have been notable advancements in mapping the structure of this field \cite{Xu:2024}. 
Consequently, studying theoretically the deflection of the charged signal by magnetic field becomes much more meaningful, aiming for their potential application in more realistic environments in forthcoming studies.

Many approaches have been employed to investigate the deflection of charged signals theoretically in both weak and strong gravitational fields in fixed spacetimes  \cite{Crisnejo:2019xtp,Jusufi:2019rcw,Xu:2021rld,Zhou:2022dze}.
One specific approach involves the application of the Gauss-Bonnet theorem  \cite{Gibbons:2008rj,Crisnejo:2019xtp,Jusufi:2019rcw,Li:2022cpu,Li:2023svb}, which is applicable in the presence of electromagnetic force. Another method utilizes a perturbative strategy that employs a broadly applicable integral transformation technique  \cite{Huang:2020trl,Xu:2021rld,Zhou:2022dze}.
This method is suitable for various stationary and axisymmetric (SAS) spacetimes, accommodating both null and timelike signals. 

In this work, we will use the perturbative method to study the effect of magnetic fields on the deflection angle of charged signals in the weak deflection limit. 
We will show that the method works in arbitrary such SAS spacetimes with quite general forms of electromagnetic potential. Moreover, the results can reach arbitrary accuracy and automatically take into account the finite distance effect of the signal source and detector. Our deflection result also reveals the relative strength of the gravitational, electrostatic, and magnetic contributions to the deflection angle in various scenarios. 

The paper is organized as follows. 
In Sec. \ref{sec:method}, the general methodology is introduced and the deflection angle to arbitrary post-Newtonian (PN) order is found. 
Sec. \ref{sec:application} demonstrates the application of this method to various scenarios, including the classical Kerr–Newman (KN) spacetime \cite{Newman:1965my,Carter:1968rr}, Kerr spacetime with a dipole magnetic field \cite{Petterson:1975sg,Takahashi:2008zh}, and a mass possessing a magnetic dipole moment \cite{Gutsunaev:1987}. The accuracy of the PN results in these cases was confirmed by comparing with the numerical result, and the effects of various signal or spacetime parameters on the deflection angle are analyzed. Sec. \ref{sec:conc} concludes the paper by summarizing the important findings and giving an outlook.
In the work, we use the natural system of units $G=c=4\pi\varepsilon_{0}=1$.

\section{Methodology in general SAS spacetimes\label{sec:method}}

The spacetime with axial symmetry is usually described using the Boyer-Lindquist coordinates $(t,r,\theta,\phi)$ as
\be
\mathrm{d}s^2=-A\mathrm{d}t^2+B\mathrm{d}t\mathrm{d}\phi+C\mathrm{d}\phi^2+D\mathrm{d}r^2+F\mathrm{d}\theta^2,
\label{eq:metricg}
\ee
in which the metric functions $A,\,B,\,C,\,D,\,F$ rely on $r$ and $\theta$ only. For a spacetime with a magnetic field, if the motion of charged particles can be kept on the equatorial plane  $\left(\theta=\frac{\pi}{2}\right)$, which we will concentrate on in this paper, there will be strong requirements for the electromagnetic potential. Here, we only study an electromagnetic field with the four-potential of the form  $\left[\mathcal{A}_t\left(r,\theta\right),0,0,\mathcal{A}_\phi\left(r,\theta\right)\right]$, which satisfies $\partial_\theta \mathcal{A}_t = \partial_\theta \mathcal{A}_\phi = 0$ on the equatorial plane.

The motion of a test particle with a mass $m$ and an electric charge $q$, and therefore the specific charge $\hat{q}\equiv q/m$, in this spacetime and magnetic field is described by the Lorentz equation \cite{Chase:1954}
\be\label{eq:Lorentz}
\frac{\dd ^2x^\rho}{\dd \tau^2}+\Gamma^{\rho}{}_{\mu\nu}\frac{\dd x^\mu}{\dd \tau}\frac{\dd x^\nu}{\dd \tau}=\hat{q}\mathcal{F} ^{\rho}{}_{\mu}\frac{\dd x^\mu}{\dd \tau},
\ee
where $\mathcal{F}_{\mu\nu}=\partial_\mu \mathcal{A}_\nu-\partial_\nu \mathcal{A}_\mu$ and $\tau$ is the proper time of the test particle. Integrating Eq. (\ref{eq:Lorentz}) on the equatorial plane, we obtain the first-order motion equations  \cite{Duan:2023gvm}
\begin{subequations}\label{eqs:motions}
\begin{align}
\dot{\phi} =&\frac{2\left(2\Lambda A-\Xi B\right)}{B^{2}+4AC},  \label{eq:phimotion}\\
\dot{t} =&\frac{2\left(\Lambda B+2\Xi C\right)}{B^{2}+4AC}, \label{eq:tmotion}\\
\dot{r}^{2} =&\frac{\left(\Xi ^2-A\right)(B^2+4AC)-\left(2\Lambda A-\Xi B\right)^2}{AD\left(B^2+4AC\right)},\label{eq:rmotion}
\end{align}
\end{subequations}
where we have defined some auxiliary symbols
$\Lambda=\hat{L}-\hat{q}\mathcal{A}_\phi,\,\Xi=\hat{E}+\hat{q}\mathcal{A}_t$
in which $\hat{E}$ and $\hat{L}$ correspond to the energy and angular momentum per unit mass of the particle respectively, and the dot  $\dot{{}}$ denotes the derivative with respect to $\tau$. 
Note that the sign of $\hat{L}$ in general is not equivalent to the sign of $\dot{\phi}$ along the entire trajectory of the particle, especially in high curvature regions. Only in the weak field limit (i.e., at spatial infinity), based on the fact that $\hat{L}\sim r^{1/2}$ and other general assumptions, we can conclude that the signs of $\hat{L}$ and $\dot{\phi}$ along that part of the trajectory are the same.

In the context of asymptotically flat spacetimes, the constants $\hat{E},\,\hat{L}$ can be correlated with the impact parameter $b$ and the asymptotic velocity $v$ of the signal, which can be observed at infinity, as
\begin{equation}\label{eq:EandL}
    \hat{E}=\frac{1}{\sqrt{1-v^2}},~\hat{L}=\frac{s_0 b v}{\sqrt{1-v^2}}=s_0 b \sqrt{\hat{E}^2-1},
\end{equation}
in which $s_0=\pm 1$ represents the direction of particle orbit in the weak field limit, with positive and negative signs corresponding to counterclockwise and clockwise directions, respectively. The relationship between $\hat{L}$ and the minimum value $r_0$ of radial coordinate can be deduced from the definition equation $\left.\dot{r}\right|_{r=r_0}=0$ using Eq. (\ref{eq:rmotion})
\begin{align}\label{eq:r0}
&2\Lambda\left(r_0\right) A(r_{0})-\Xi\left(r_0\right)B(r_{0})=\nn\\
&s_1 \sqrt{\left[4A\left(r_{0}\right)C\left(r_{0}\right)+B\left(r_{0}\right)^{2}\right]\left[\Xi\left(r_0\right)^{2}-A\left(r_{0}\right)\right]},
\end{align}
where $s_1=\pm 1$ indicates the specific direction in which the particle orbits the central celestial body at $r_0$. As previously mentioned, in the case of weak deflection, $s_1$ is equivalent to $s_0$, and hence we will denote them together by $s$. Consequently, the connection between the impact parameter $b$ and the distance $r_0$ can be expressed from Eq. (\ref{eq:EandL}) as follows
\be\label{eq:p}
\begin{aligned}
\frac1b& =\frac{2 A(r_{0})\left[\sqrt{\hat{E}^{2}-1}-s\hat{q}\mathcal{A}_\phi(r_{0})/b \right]-s B(r_{0})\Xi\left(r_0\right)/b}{\sqrt{\left[4A\left(r_{0}\right)C\left(r_{0}\right)+B\left(r_{0}\right)^{2}\right]\left[\Xi\left(r_0\right)^{2}-A\left(r_{0}\right)\right]}} \\
&\equiv p\left(\frac{1}{r_{0}}\right),
\end{aligned}
\ee
where for later usage we have defined the right-hand side as a function $p$ of $1/r_0$ and we will denote the inverse function of $p(x)$ as $h(x)$.

From Eqs. (\ref{eq:phimotion}), (\ref{eq:rmotion}) and (\ref{eq:r0}), the deflection angle $\Delta \phi$ of a signal originating from a source located at radius $r_s$ to a detector at $r_d$ can be expressed as  \cite{Huang:2020trl}
\begin{equation}\label{eq:dphi_raw}
\begin{aligned}
\Delta\phi=& \left(\int_{r_0}^{r_s}+\int_{r_0}^{r_d}\right) \frac{\sqrt{AD}}{\sqrt{B^2+4AC}}\\
&\times \frac{2\left(2\Lambda A-\Xi B\right) \mathrm{d}r}{\sqrt{\left(\Xi^2- A\right)(B^2+4AC)-\left(2\Lambda A-\Xi B\right) ^2}}.
\end{aligned}
\end{equation}
This integral, however, in general can not be carried out to find a closed form. Therefore, we will use the PN approach to first expand it in the large orbit limit and then integrate the resultant series. 
To this end, we make a change of variables first from $r$ to $u$ which is linked by 
the function $h$ defined after Eq. \eqref{eq:p}, i.e., 
\be\label{eq:qdef}
\frac{1}{r}=h\left(\frac{u}{b}\right),
\ee
so that $\Delta\phi$ becomes 
\begin{align}\label{eq:integrand2}
\Delta\phi
=& \left(\int_{\sin\delta_{s}}^{1}+\int_{\sin\delta_{d}}^{1}\right)\sqrt{\frac{A(1/h)D(1/h)}{A(1/h)C(1/h)+B(1/h)^{2}/4}}\nn\\
&\times \frac{1}{p^{\prime}\left(h\right)h^{2}}\frac{u}{b}\frac{\mathrm{d}u}{\sqrt{1-u^{2}}},
\end{align}
where 
\begin{equation}\label{eq:apperentangle}
\delta_{s,d}=\arcsin\left[b\cdot p\left(\frac{1}{r_{s,d}}\right)\right]
\end{equation}
are apparent angles associated with the impact parameter and the radial coordinates of the source and detector \cite{Huang:2020trl}.
Note here the symbols $h$ should be considered as the function $h(u/b)$ so that the entire integrand is about the variable $u$. The primary purpose of this specific change of variable is to make the integral range within $(0,1)$ so that the latter small $u$ expansion of the integrand, i.e., the PN approximation, can be carried out smoothly. 

In the integrand of Eq. (\ref{eq:integrand2}), the factor $1/{\sqrt{1-u^{2}}}$ is simple and can be kept without expanding. We will then denote the rest of the integrand as $y(u,b)$ and then expand it for small $u$
\be\label{eq:y_expansion}
\begin{aligned}
    y\left(u,b\right)=& \sqrt{\frac{A(1/h)D(1/h)}{A(1/h)C(1/h)+B(1/h)^{2}/4}}\frac{1}{p^{\prime}\left(h\right)h^{2}}\frac{u}{b}\\
    =&\sum_{n=0}^{\infty}y_{n}(b)\left(\frac ub\right)^{n}.
\end{aligned}
\ee
Note that once the metric and four-potential functions are known, this expansion becomes immediately feasible; therefore, the coefficients $y_n$ are related to these functions, as well as the other test particle parameters including $\hat{q},\,b,\,v$ etc. We will postpone giving the exact forms of the first few $y_n$ until Eq. \eqref{eq:ynres}.
The expansion \eqref{eq:y_expansion} transforms the integral (\ref{eq:integrand2}) into a series of the form
\be\label{eq:phiOny}
\Delta\phi=\left(\int_{\sin \delta_s}^{1}+\int_{\sin \delta_d}^{1}\right) \sum_{n=0}^{\infty}\frac{y_n(b)}{b^n} \frac{u^n}{\sqrt{1-u^{2}}}\mathrm{d}u.
\ee
Because integrals of the following form can always be carried out explicitly
\be\label{eq:ln}
\left(\int_{\sin \delta_s}^{1}+\int_{\sin \delta_d}^{1}\right)  \frac{u^n}{\sqrt{1-u^{2}}}\mathrm{d}u \equiv l_n(\delta_s,\delta_d)~~(n=0,1,2,\cdots),
\ee
where we used $l_n$ to denote the results with their explicit forms shown in Appendix \ref{sec:ln}, the integrability of the deflection \eqref{eq:phiOny}, and consequently the effectiveness of our PN method, are also guaranteed. Note that the finite distance effect of the signal source and detector lies in these $l_n$ through $\delta_s$ and $\delta_d$. 
Substituting Eq. \eqref{eq:ln} into Eq. \eqref{eq:phiOny}, we found 
\be\label{eq:dphiinlshort}
\Delta\phi=\sum_{n=0}^{\infty}\frac{y_n(b)l_n(\delta_s,\delta_d)}{b^n}.
\ee

This is only a quasi-series of the impact parameter $b$ yet because there is still weak $b$ dependence in the coefficients $y_n$ and $l_n$. In principle, one can further expand the latter two into series of $b$ too and consequently, the entire deflection becomes a true power series of the impact parameter. However, since the $l_n$ depend on $b$ only very weakly and it contains positive powers of $b/r_{s,d}$ (see Appendix \ref{sec:ln}), we will not expand $l_n$ but only $y_n(b)$ into powers of $b$. Latter study further shows that the resultant series can always be completely split into two categories, 
the neutral part with coefficients denoted as $\beta_n$ and the electromagnetic part with coefficient $\gamma_n$, 
\be\label{eq:phiOnZ}
\Delta\phi=s \sum_{n=0}^{\infty}\frac{\beta_n(l_i,v)+\gamma_n(l_i,v)}{b^n},
\ee
where both kinds of coefficients are functions of $r_s,\,r_d$ through $l_i$ and particle velocity $v$ and other spacetime parameters. These two parts have no overlapping or coupling, indicating that in the deflection angle computation, the gravitational and electromagnetic contributions are intrinsically different and can be completely separated. 

To derive the exact forms of $\beta_n$ and $\gamma_n$ in Eq. \eqref{eq:phiOnZ}, we will need to obtain the $y_n$ in Eq. \eqref{eq:dphiinlshort} first. We will assume that the metric and four-potential functions admit the following asymptotic expansions
\be\label{eq:metricCoe}
\begin{aligned}
    &A(r)=1+\sum_{n=1}^\infty\frac{a_n}{r^n},
    \quad && B(r)=\sum_{n=1}^\infty\frac{b_n}{r^n},\\
    &\frac{C(r)}{r^2}=1+\sum_{n=1}^\infty\frac{c_n}{r^n},
    \quad && D(r)=1+\sum_{n=1}^\infty\frac{d_n}{r^n},\\
    &\mathcal{A}_t(r)=\sum_{n=1}^\infty\frac{\mathfrak{q}_{0n}}{r^n},
    \quad && \mathcal{A}_\phi(r)=\sum_{n=1}^\infty\frac{\mathfrak{q}_{3n}}{r^n},
\end{aligned}
\ee
which implies that we focus on asymptotically flat spacetimes in this work.
The expansion of the electric potential $\mathcal{A}_t$ starts from order $r^{-1}$ just as in the case of the Reissner-Nordstr\"om (RN) and many other charged spacetimes. For the magnetic potential, we also assume its asymptotic form starts from the $r^{-1}$ order, like a magnetic dipole. Magnetic fields with stronger asymptotics, such as a uniform magnetic field with asymptotics of order $r^2$, can not be dealt by the PN method because the Lorentz force in that case will dominate in the asymptotic region, and therefore the motion is not (post-)Newtonian. 

With the coefficients in Eq. (\ref{eq:metricCoe}), the first few terms in Eq. (\ref{eq:phiOny}) can be deduced after using Eqs. (\ref{eq:qdef}) and (\ref{eq:y_expansion}) as
\be \label{eq:ynres}
\begin{aligned} 
    y_0=& s,\\
    y_1=& s \left(\frac{2  \hat{q} \mathfrak{q}_{01} \sqrt{1-v^2}-a_1 }{2  v^2}+\frac{d_1}{2}\right)\\
    &+\frac{1}{b v}\left(\frac{b_1}{2 }+\hat{q} \mathfrak{q}_{31}   \sqrt{1-v^2}\right).
\end{aligned}
\ee
Higher order $y_n$'s can also be obtained easily but are too tedious to present here. Substituting these $y_n$'s into Eq. \eqref{eq:dphiinlshort} and expanding again, we found the first few $\beta_n$ and $\gamma_n$ as
\begin{subequations}\label{eqs:zn}
\begin{align}
    \beta_{0}=&  l_0,\\
    \beta_{1}=&  \left(\frac{d_1}{2}-\frac{a_1}{2v^2}\right)l_1,\label{eq:b1coeff}\\
    \beta_{2}=& \left[\frac{c_2+d_2}{2}-\frac{(c_1-d_1)^2}{8}+\frac{2a_1^2-2a_2-a_1\left(c_1+d_1\right)}{2v^2}\right]l_2\nn\\
    &+s \frac{b_1}{2v}l_1,\\
    \gamma_{0}=& 0,\\
    \gamma_{1}=&  \frac{ \hat{q} \mathfrak{q}_{01}  \sqrt{1-v^2} }{  v^2}l_1,\label{eq:g1coeff}\\
    \gamma_{2}=&\hat{q}\frac{\sqrt{1-v^2}}{v}\left[s \mathfrak{q}_{31}l_1+ l_2\frac{\mathfrak{q}_{01} \left(-2 a_1+c_1+d_1\right)+2 \mathfrak{q}_{02} }{v}\right]\nn\\
    &+ \hat{q}^2  l_2 \mathfrak{q}_{01}^2 \left(\frac{1-v^2}{v^2}\right).\label{eq:g2coeff}
\end{align}
\end{subequations}
Higher order coefficients can be obtained without much difficulty but are too  long to present here. 

It is important to recognize the following points about this result for better evaluation of different kinds of contributions to the deflection. The first is about the leading 
$(1/b)^1$ orders from which the gravitational and electromagnetic part contribute. It is observed from Eqs. \eqref{eq:b1coeff} and \eqref{eq:g1coeff} that the gravitational part of the deflection appears from the $(1/b)^1$ order while the earliest possible order of the electromagnetic part is also the $(1/b)^1$ order, provided the $\mathfrak{q}_{01}$ coefficient of the electrostatic potential $\mathcal{A}_t$ is nonzero. 
This last requirement is not necessarily true if the spacetime itself is not charged. For those spacetimes without their own charge, then their electromagnetic contribution to the deflection will appear from higher orders. The second point is that compared to the gravitational part, the electromagnetic contribution terms, labeled by powers of $\hat{q}$, are always multiplied by the reciprocal Lorentz factor $\sqrt{1-v^2}$ or its positive powers. Mathematically, this tracks back to Eq. \eqref{eq:p}. When factoring out $\sqrt{\hat{E}^2-1}=v/\sqrt{1-v^2}$ in the numerator of this equation, terms proportional to $\hat{q}\mathcal{A}_\phi$ and $\hat{q}\mathcal{A}_t$ or their powers
will automatically contain a factor of $\sqrt{1-v^2}$ or its integer powers. This origin implies that the electromagnetic part of the deflection is strongly suppressed for relativistic particles, in agreement with what is observed in Ref. \cite{Xu:2021rld,Li:2023svb}.  The third point is that besides the above suppression, the electromagnetic contribution can also be enhanced, by factor $\hat{q}$ and its powers. Indeed, the specific charges of most particles, such as electrons, protons or light nuclei, are extremely large compared to the charge/mass ratio of the gravitational center. 
The above two points are fundamental differences between the bending due to gravity and electromagnetic interaction. A fourth point is about the difference between the leading electrostatic contribution given in Eq. \eqref{eq:g1coeff} and the magnetostatic contribution
in Eq. \eqref{eq:g2coeff}.
The latter starts from the order $1/b^2$ (term proportional to $\mathfrak{q}_{31}$) while the former could in principle start from one order earlier, again provided $\mathfrak{q}_{01}$ is nonzero. However, as we will see from Subsec. \ref{subsec:kd}, the magnetic contribution might be as large or even larger than the electric one if the central charge is absent/small, but the magnetic field is strong. 

Although result \eqref{eq:phiOnZ} contains the finite distance effect of the source and detector, the infinite $r_{s,d}$ form of this deflection is also useful for others' reference. Taking this limit in Eqs. \eqref{eqs:zn} so that the $l_i$ reduce to their constant values according to Appendix \ref{sec:ln}, the deflection becomes much simpler
\begin{widetext}
\be\label{eq:Deltaphi_on_b}
\begin{aligned}
    \Delta\phi=&s \left[\pi +  \left(\frac{2 \mathfrak{q}_{01}  \hat{q} \sqrt{1-v^2}-a_1 }{  v^2}+d_1\right)\frac{1}{b}+ \left\{\frac{\pi }{2} \left[\frac{c_2+d_2}{2}-\frac{(c_1-d_1)^2}{8}+\frac{2a_1^2-2a_2-a_1\left(c_1+d_1\right)}{2v^2}\right]\right.\right.\\
    &\left.\left.+\frac{s b_1}{v}+\hat{q}\frac{\sqrt{1-v^2}}{v}\left[2 s \mathfrak{q}_{31}+ \frac{\pi}{2}\frac{\mathfrak{q}_{01} \left(-2 a_1+c_1+d_1\right)+2 \mathfrak{q}_{02} }{v}\right]+\frac{\pi }{2} \hat{q}^2   \mathfrak{q}_{01}^2 \left(\frac{1-v^2}{v^2}\right)\right\}\frac{1}{b^2} \right]+\mathcal{O}\left[\frac{1}{b}\right]^3.
\end{aligned}
\ee
\end{widetext}

\section{Application}\label{sec:application}

In this section, we apply the above method and results to some important magnetized spacetimes, and investigate the effects of electromagnetic interaction and other spacetime and particle parameters on the deflection of charged particles.

\subsection{KN spacetime\label{subsec:KN}}

The KN spacetime is one of the most important charged SAS spacetimes in the literature. The metric functions and the  electromagnetic potential in its equatorial plane $\theta=\pi/2$ are  \cite{Carter:1968rr}
\begin{subequations}\label{eqs:KNMetric}
\begin{align}
A\left(r\right)&=\Delta_-/r^2,\\
B\left(r\right)&=-2a(2Mr-Q^2)/r^2,\\
C\left(r\right)&=r^2+a^2\Delta_+/r^2,\\
D\left(r\right)&=r^2/(a^2+\Delta_-),\\
\mathcal{A}_{\mu}\left(r\right)&=\left(-\frac{Q}{r},0,0,\frac{a Q \eta }{r}\right),\label{eq:kn4p}
\end{align}
\end{subequations}
where $\Delta_{\pm}(r)=r^{2}\pm(2Mr-Q^{2})$ and $M,~Q,~a$ denote the mass, electric charge, and angular momentum per unit mass respectively. Note that in the $\mathcal{A}_\phi$ component, we have introduced a flag symbol $\eta$ to distinguish in the following steps whether certain terms containing $\hat{q},\,Q$ and $a$ are from the magnetic interaction. In other words, the terms containing $\eta$ also contain magnetic field contribution. Eventually, in numerical computations or figure plotting, it will be set to one.  

Expanding these functions as Eqs. (\ref{eq:metricCoe}) and  substituting the expansion coefficients into Eqs. (\ref{eqs:zn}), it becomes routine to derive $\Delta\phi_{\text{KN}}$ using Eq. (\ref{eq:phiOnZ}) for the KN spacetime
\begin{align}
    \Delta \phi_{\text{KN}}=s \sum_{i=0}^{\infty} \frac{\beta_{i}^{\text{KN}}+\gamma_i^{\text{KN}}}{b^i}  ,\label{eq:knd}
\end{align}
with the first few coefficients $\beta^{\mathrm{KN}}_i$ for the neutral part as
\begin{subequations}
\label{eq:KN_Neu}
    \begin{align}
    \beta_{0}^{\text{KN}}=&  l_0,\label{eq:NK_Neu_0}\\
    \beta_{1}^{\text{KN}}=&  l_1 M \left(1+\frac{1}{v^2}\right),\label{eq:KN_Neu1}\\
    \beta_{2}^{\text{KN}}=& -\frac{2 s l_1 a M}{v}+ l_2 \left[\frac{3 M^2}{2}+\frac{6 M^2}{v^2}-Q^2\left( \frac{1}{2}+\frac{1}{v^2}\right)\right], \label{eq:KN_Neu2}
\end{align}
\end{subequations}
and coefficients $\gamma^{\mathrm{KN}}_i$ for the electromagnetic part as
\begin{subequations}\label{eq:KN_EM}
\begin{align}
    \gamma_{0}^{\text{KN}}=&0,\\
    \gamma_{1}^{\text{KN}}=&  -l_1 \hat{q}Q\frac{\sqrt{1-v^2}}{v^2},\label{eq:KN_EM_1}\\    \gamma_{2}^{\text{KN}}=&\hat{q} Q \sqrt{1-v^2} (\frac{s a \eta  l_1 }{v}-\frac{6 l_2 M }{v^2})-l_2 \hat{q}^2 Q^2 \left(1-\frac{1}{v^2}\right),\label{eq:KN_EM_2}\\
    \eta=&1.\nonumber
\end{align}
\end{subequations}
The neutral part up to order $2$ is found to be consistent with results in Ref.  \cite{Huang:2020trl}.
Besides the fact pointed out after Eq. \eqref{eqs:zn} that the magnetic contribution is one order higher than the electric one, we also note that the magnetic contribution is scaled by the spacetime spin, which is a consequence of the form of the magnetic four-potential in Eq. \eqref{eq:kn4p}. 

\begin{figure}[htp!]
\centering
\includegraphics[width=0.45\textwidth]{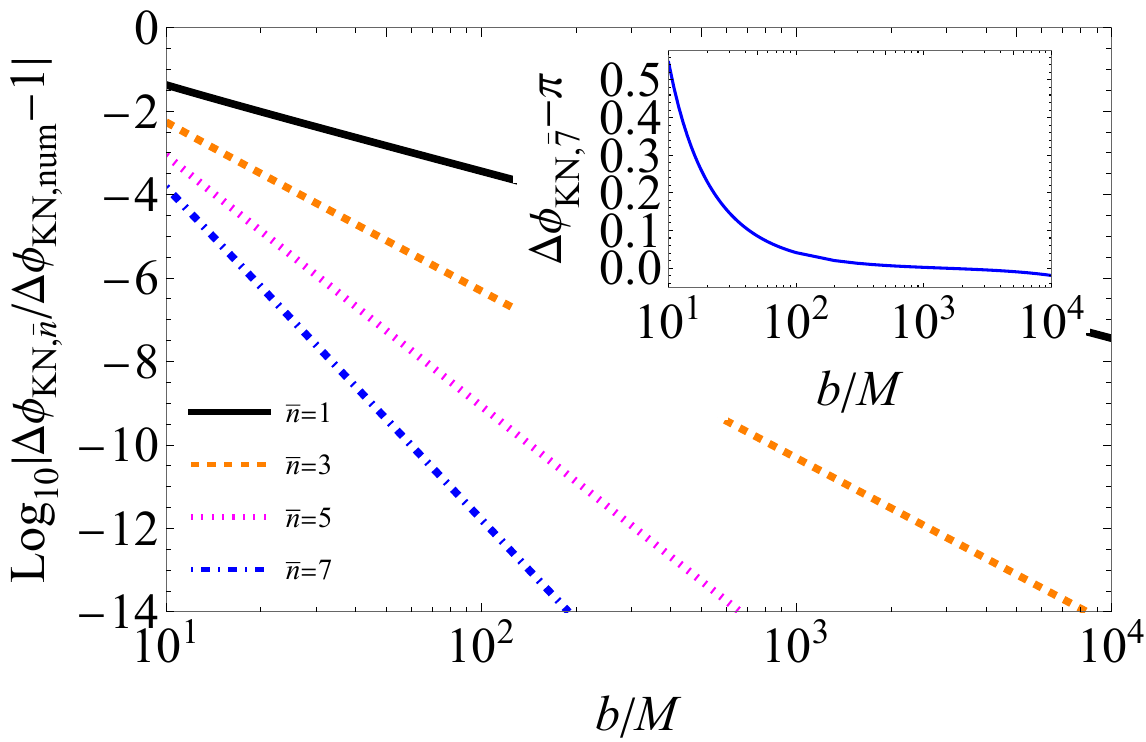}
\caption{The perturbative $\Delta\phi_{\text{KN}, \bar{n}}$ and its difference from the numerical result $\Delta\phi_{\text{KN},\text{num}}$ as functions of $b/M$ from $10^1$ to $10^4$. Other parameters used are $Q=M/2,~a=M/3,~\hat{q}=1/10,~s=1,~v=1-10^{-2}$ and $r_s=r_d=10^6M$. }\label{fig:KN_Err}
\end{figure}

To check the correctness of $\Delta\phi_{\text{KN}}$ in \eqref{eq:knd} and study the magnetic contribution, we can define a truncated deflection $\Delta\phi_{\text{KN},\bar n}$ consisting of terms in Eq. \eqref{eq:knd} up to the $\bar{n}$-th order, i.e.,
\begin{align}
\Delta\phi_{\text{KN},\bar n}=s \sum_{i=0}^{\bar n} \frac{\beta_{i}^{\text{KN}}+\gamma_i^{\text{KN}}}{b^i}. \label{eq:knnbar}
\end{align}
Moreover, since the gravitational and pure electric effects on the deflection have been well studied previously  \cite{Jia:2020xbc,Huang:2020trl,Xu:2021rld}, in this work, we will concentrate on the effect of the magnetic field. To quantitatively do so, we define a magnetic contribution part of the deflection angle as
\be\label{eq:KN_zn_eta}
\Delta\phi^{\text{Mag}}_{\text{KN},\bar{n}}=\left.\Delta\phi_{\text{KN},\bar n}\right|_{\eta=1}-\left.\Delta\phi_{\text{KN},\bar n}\right|_{\eta=0}.
\ee
In Fig. \ref{fig:KN_Err},
we first compare the truncated deflection $\Delta\phi_{\text{KN},\bar n}$ with the deflection angle obtained using numerical integration, $\Delta\phi_{\text{KN,num}}$. Due to the high accuracy of numerical integration, $\Delta\phi_{\text{KN,num}}$ can be considered as the true deflection. The figure shows that for a fixed $b$ as the truncation order $\bar n$ increases, or for a fixed $\bar n$ as $b$ increases, the perturbative deflection angle converges exponentially toward the true value of the deflection. The former decreasing indicates the correctness of our perturbative result \eqref{eq:knd} and the latter decreasing reflects the fact that the result is a series of $1/b$. As the impact parameter $b$ increases, we see that the deflection itself (see the insert) decreases monotonically, roughly as $1/b$, as dictated by Eqs. \eqref{eq:KN_Neu1} and \eqref{eq:KN_EM_1}.  

Now with the correctness of the deflection \eqref{eq:knd} confirmed, we can use it to study the effect of various parameters on the deflection. However, before that, it is worthy to analyze the degeneracy of the parameter space spanned by $(a,\,Q,\,q,\,s)$ first. 
Observing from Eqs. \eqref{eq:KN_Neu} and \eqref{eq:KN_EM} that altering the signs of $a$ and $s$ (or $\hat{q}$ and $Q$) simultaneously will not affect the size of either the neutral part or the electromagnetic part of the deflection, but only the sign of $\Delta\phi_\text{KN}$.
The invariance of the size of the deflection angle under simultaneous sign change of $s$ and $a$ can be simply understood as follows. When the magnetic field undergoes a directional flip due to the sign alteration of $a$, it is sufficient to modify the rotation direction $s$ of the signal to maintain the direction (inward or outward) of the Lorentz force exerted on the particle and consequently the deflection angle size. On the other hand, although a sign change of $Q$ also flips the direction of the magnetic field, one can not keep the size of the deflection angle by switching the signal rotation direction. The reason is that in this case the spacetime rotation direction $a$ is not changed and it will alter its contribution to the deflection angle through the gravitational channel, and consequently changes the total deflection size. In contrast, the effect of the sign change of $Q$ can be exactly compensated if $\hat{q}$ changes sign. This ensures that the Lorentz force is not changed, as well as the electric force and gravitational interaction. 
In summary, the above degeneracies allow us to concentrate on the positive $a$ and $Q$ while maintaining general $s$ and $\hat{q}$ when studying effects of these parameters.
Note from Eq. \eqref{eq:kn4p} that the above sign choice means that the direction of the magnetic field is also fixed, in this case downward from the equatorial plane, i.e., with a nonzero component $B_\theta>0$. Similarly, the direction of the electric field is fixed outward on the equatorial plane, i.e., with a component $E_r>0$.

\begin{figure}[htp!]
\centering
\includegraphics[width=0.45\textwidth]{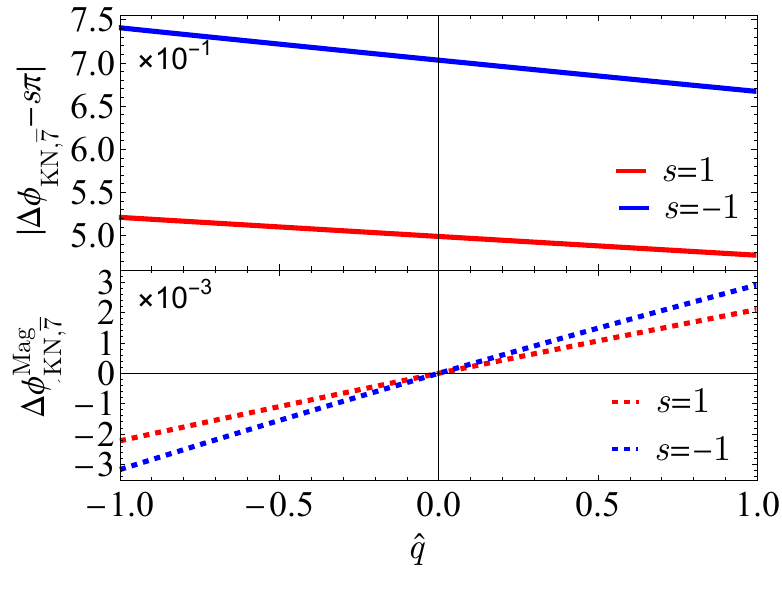}
\caption{$\Delta\phi_{\text{KN}, \bar{7}}$ and the contribution $\Delta\phi^{\text{Mag}}_{\text{KN},\bar{7}}$ of the magnetic field as functions of $\hat{q}$ with other parameters $Q=M/2,~a=9/10 M,~v=1-10^{-2}, ~r_s=r_d=10^5M,~b=10M$. }
\label{fig:KNonq}
\end{figure}

In Fig. \ref{fig:KNonq}, we illustrate the effect of the specific charge $\hat{q}$ on the non-trivial part (order $1$ and above) of the deflection with chosen parameters $Q=M/2$ and $a=9M/10$.
For small $|\hat{q}|$ such that the electromagnetic force is considerably weaker than the gravitational one, an increase in $\hat{q}$ (around $0$) will lead to a greater attenuation of the gravitational attraction by the electrostatic interaction. This results in a reduction of $|\Delta\phi_\text{KN}|$, as can be seen from the comparison of Eq. \eqref{eq:KN_Neu1} and \eqref{eq:KN_EM_1} and confirmed by the two solid lines in Fig. \ref{fig:KNonq}. For much larger $|\hat{q}|$ so that the electrostatic interaction is much stronger than the gravitational one, the signal trajectory can even bend away from the center if $\hat{q}Q>0$. However, the effect of the pure electric interaction has been well studied in Ref. \cite{Xu:2021rld} and we will not pursue along this line. 

Since the only nonzero component of the magnetic field $B_\theta$ is greater than zero because both $a$ and $Q$ are chosen positive, when a signal is positively charged ($\hat{q}>0$), its Lorentz force will be attractive (or repulsive) if it is counterclockwise rotating (or clockwise rotating) so that the magnetic contribution to the deflection is positive. 
The opposite thing happens when $\hat{q}$ changes sign. This agrees with the dashed lines in Fig. \ref{fig:KNonq}.
The difference in the slopes of the two dashed lines is due to different sign contributions of the spin $a$ through the gravitational channel. Finally, we note a small deviation of the solid lines from perfect linearity, which is due to higher-order corrections, such as the $\hat{q}^2$ term in Eqs. (\ref{eq:KN_EM}).

\begin{figure}[htp!]
\centering
\subfigure[]{\includegraphics[width=0.45\textwidth]{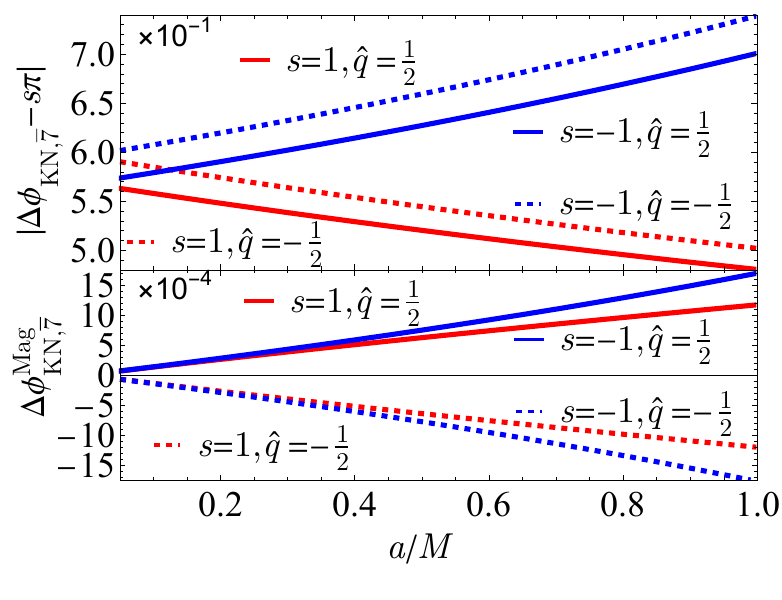}\label{fig:KN_on_a}}
\subfigure[]{\includegraphics[width=0.45\textwidth]{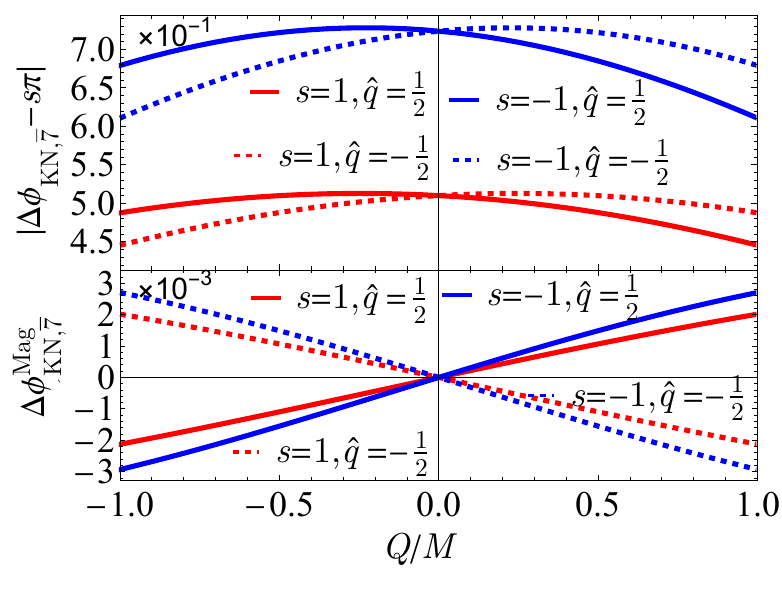}\label{fig:KN_on_Q}}
\caption{$\Delta\phi_{\text{KN}, \bar{7}}$ and its magnetic part $\Delta\phi_{\text{KN}, \bar{7}}^{\text{Mag}}$ as functions of $a$ and $Q$. Other parameters used are $Q=M/2,~a=9/10 M,~v=1-10^{-2}, ~r_s=r_d=10^5M,~b=10M$, except for the running one. \label{fig:KN_on_aandq}}
\end{figure}

In Fig. \ref{fig:KN_on_aandq}, the total deflection angle and its magnetic part are plotted as functions of $a$ and $Q$ respectively, for different signs of $s$ and $\hat{q}$. For the total deflection in the top panel of Fig. \ref{fig:KN_on_aandq} (a), we see that as $a$ increases, prograde moving particles with $s=1$ represented by red curves (or retrograde moving particles with $s=-1$ marked by blue curves) will have a decreasing (or increasing) deflection angle, regardless of the charge $\hat{q}$'s sign. This agrees with the effect of $a$ on the deflection of neutral particles \cite{Jia:2020xbc}, suggesting the effect is through the gravitational channel, i.e., the first term in Eq. \eqref{eq:KN_Neu2}. Since in this panel $Q=M/2>0$, the positive (or negative) charges experience a repulsive (or attractive) electrostatic force, its deflection is slightly smaller (or larger) than that of negative (or positive) charges. This is confirmed since the solid curves are always below the corresponding dashed lines of the same color, which is also dictated by Eq. \eqref{eq:KN_EM_1}. If one inspects more carefully the gap between the two red (or blue) curves, it is seen that the gap size decreases (or grows) slightly as $a$ increases. These gap changes are due to the fact that as $a$ increases, the Lorentz force that the magnetic field asserted on the charges of two different signs also increases, in accord with Eq. \eqref{eq:KN_EM_2}. 

The effect of the magnetic force can be more clearly seen from the bottom panel of Fig. \ref{fig:KN_on_aandq} (a), which shows specifically $\Delta \phi_\text{KN}^\text{Mag}$. The first thing we notice from this is that the magnetic contribution in general is much smaller than the total deflection angle $\Delta \phi_\text{KN,{$\bar 7$}}$ in the top panel, which is in accord with the observation made below Eqs. \eqref{eqs:zn} that the magnetic contribution only appears from the order $1/b^2$. The second general point is that all trends of these magnetic contributions as $a$ increase are in agreement with the effect of the Lorentz force on the charges, as seen in the first term of Eq. \eqref{eq:KN_EM_2}.  Moreover, we also see that when $\hat{q}$ switches sign, the magnetic contribution becomes exactly the opposite, i.e., from the solid curve to the same colored dashed curve, which is due to the sign change of the Lorentz force. However, when the rotation direction $s$ changes sign, not only does the Lorentz force change direction but the gravitational part is also modified through the first term in Eq. \eqref{eq:KN_Neu2}. Therefore, the two solid (or dashed) lines in the top panel do not completely overlap. 

In Fig. \ref{fig:KN_on_Q}, we plot the above quantities for different $Q$. 
Our previous analysis of the parameter space degeneracy suggests that it is sufficient to concentrate only on positive $Q$'s. Nevertheless, we have incorporated in this plot the negative $Q$ to provide a comprehensive depiction of the parabolic-like curves.
The parabolic shape of these curves, in general, can be attributed to the gravitational impact of $Q^2$, whereas the electromagnetic force dictates the shift of the extreme point of the parabola to the left/right. When $\hat{q}=0$, i.e., the electromagnetic effect on the signal is absent, the deflection angles are symmetric about the $Q=0$ axis and decrease as $|Q|$ increases, in agreement with the result in the RN spacetime \cite{Pang:2018jpm}. For positively charged particles traveling in a counterclockwise direction, a slightly negative $Q$ will generate a static electric attraction that is greater than the influence of the magnetic field, thereby increasing the total deflection angle and finally resulting in a shift of the maximum point towards the negative region of $Q$, as seen from the red solid line in the top panel of Fig.  \ref{fig:KN_on_aandq} (b). 
For small values of $Q$ and $a$, the term containing $\eta$ in Eq. (\ref{eq:KN_EM}) represents the primary contribution from the magnetic field to the deflection angle. This term exhibits similar dependence on both $Q$ and $a$. Consequently, the features of the lines plotted in the bottom panel Fig.  \ref{fig:KN_on_aandq} (b) and \ref{fig:KN_on_aandq} (a) are similar.

\subsection{Kerr spacetime with a dipole magnetic field \label{subsec:kd}}

Besides the magnetic field associated with a rotating charged spacetime such as the one studied in the above subsection, people often add a magnetic field by hand into the spacetime to mimic certain astronomical scenarios. One popular such magnetic field is the dipolar one presented in Ref. \cite{Petterson:1975sg,Takahashi:2008zh}. In this subsection, we study the deflection in Kerr spacetime with this magnetic field. We emphasize that such magnetic fields are usually assumed to be so weak that the background spacetime structure is
not altered by them.

The electromagnetic potential for a dipole magnetic field in the Kerr metric  \cite{Kerr:1963ud}
\be
\begin{aligned}\dd s^{2}=&-\left(1-\frac{2Mr}{\Sigma}\right)\dd t^{2}-\frac{4Mra\sin^{2}\theta}{\Sigma}\dd t \dd\phi+\frac{\Sigma}{\Delta}\dd r^{2} \\
&+\Sigma \dd\theta^{2}+\frac{\left(r^{2}+a^{2}\right)^{2}-\Delta a^{2}\sin^{2}\theta}{\Sigma}\sin^{2}\theta \dd\phi^{2},
\end{aligned}
\ee
where $\Sigma=r^2+a^2\cos^2\theta,\quad\Delta=r^2-2Mr+a^2$, can be written as  \cite{Petterson:1975sg,Takahashi:2008zh}
\begin{widetext}
\begin{subequations}
\label{eqs:LiPotential}
\begin{align}
\mathcal{A}_t= & -\frac{3 a \mu}{2 \Sigma \zeta^2}\left\{\left[r\left(r-M\right)+\left(a^2-M r\right) \cos^2 \theta\right]\times \frac{1}{2 \zeta} \ln \left(\frac{r-M+\zeta}{r-M-\zeta}\right)-\left(r-M \cos ^2 \theta\right)\right\} \nonumber\\
=&-\frac{a \mu }{2 r^3}-\frac{a \mu  M}{r^4}+\mathcal{O}\left[\frac{1}{r}\right]^5,\label{eq:atexp}\\
\mathcal{A}_\phi= & -\eta \frac{3 \mu \sin ^2 \theta}{4 \Sigma \zeta^2}\left\{\left(r-M\right) a^2 \cos ^2 \theta+r\left(r^2+M r+2 a^2\right)\right. \nonumber\\
&\left. -\frac{1}{2 \zeta}\left[r\left(r^3-2 M a^2+a^2 r\right)+\Delta a^2 \cos ^2 \theta\right]\times \ln \left(\frac{r-M+\zeta}{r-M-\zeta}\right)\right\}\nonumber\\
=&\eta\left[\frac{\mu }{r}+\frac{3 \mu  M}{2 r^2}+\frac{\mu  \left(a^2+24 M^2\right)}{10 r^3}-\frac{\mu  M \left(a^2-8 M^2\right)}{2 r^4}\right]+\mathcal{O}\left[\frac{1}{r}\right]^5, \label{eq:aphiexp}
\end{align}
\end{subequations}  
\end{widetext}
in which $M,~a,~\mu$ denote the mass, spin, and magnetic moment respectively. $\zeta=\sqrt{M^{2}-a^{2}}$ and $\eta$ is again a flag symbol marking the contribution of the magnetic field in later steps. Eventually, $\eta=1$ will be set.

In this spacetime, the deflection angle in the weak deflection limit can also be split into two categories, the neutral (or pure gravitational) part and the electromagnetic part. The neutral part clearly should be obtainable from that of the KN spacetime with $Q$ set to zero. However, the effect of the magnetic field in this case will be different from that studied in the KN case in the previous subsection. Here, the magnetic field at large radius resembles that of a magnetic dipole described by the magnetic moment $\mu$, whereas the electric field arises from the current loop rotating around the center. Given that the series expansion of the electric potential $\mathcal{A}_t$ begins from order $1/r^3$, the electrostatic impact is significantly diminished compared to that of the spacetime charge in KN spacetime, which starts from the $1/r$ order. In contrast, the magnetic field here still appears from the order $1/b^2$ as seen from Eq. \eqref{eq:aphiexp}. This order examination establishes the fact that the magnetic field becomes the dominant factor in electromagnetic interactions in the current situation. 

By substituting the expansion coefficients of Eq. \eqref{eqs:LiPotential} into Eqs. (\ref{eqs:zn}), it is not difficult to calculate the coefficients of the deflection angle in the current case,
\begin{align}
    \Delta \phi_{\text{KD}}=s \sum_{i=0}^{\infty} \frac{\beta_{i}^{\text{KD}}+\gamma_{i}^{\text{KD}}}{b^i}  ,
    \label{eq:kddef}
\end{align}
where the initial few orders are
\begin{subequations}\label{eqs:LDS_zn_Neu}
\begin{align}
    \beta_{0}^{\text{KD}}=&  l_0,\\
    \beta_{1}^{\text{KD}}=&  l_1 M \left(1+\frac{1}{v^2}\right),\\
    \beta_{2}^{\text{KD}}=& -\frac{2 s l_1 a M}{v}+ l_2 M^2 \left(\frac{3 }{2}+\frac{6 }{v^2}\right),\label{eq:LDS_Z2_Neu}\\
    \beta_{3}^{\text{KD}}=&-\frac{4 s l_2 a M^2}{v}\left(3+\frac{2}{v^2}\right) +  l_3 M \left[ \frac{3 a^2}{2}\left(1+\frac{1}{v^2}\right) \right.\nn\\
    &\left.+\frac{M^2}{2}\left( 5+\frac{45}{v^2}+\frac{15}{v^4}-\frac{1}{v^6}\right) \right] ,\\
    \beta_{4}^{\text{KD}}=& -\frac{3 s l_3 a M}{v}\left[a^2+3M^2\left(5+\frac{10}{v^2}+\frac{1}{v^4}\right)\right]\nn\\
    &+\left[\frac{12l_2 a^2 M^2}{v^2}+ l_4 a^2 M^2 \left(15 + \frac{40}{v^2}+\frac{8}{v^4}\right)\right.\nn\\
    &\left. +l_4 M^4 \left(\frac{35}{8} + \frac{70}{v^2}  + \frac{70}{v^4}\right) \right],
\end{align}
\end{subequations}
\begin{subequations}\label{eqs:LDS_zn_EM} 
\begin{align}
    \gamma_{0}^{\text{KD}}=&0,\\
    \gamma_{1}^{\text{KD}}=&0,\\
    \gamma_{2}^{\text{KD}}=&\eta \frac{s \hat{q}\mu \sqrt{1-v^2}}{v}l_1,\label{eq:LDS_z2_EM}\\
    \gamma_{3}^{\text{KD}}=& \frac{\hat{q} \mu \sqrt{1-v^2}}{v} \left[ -\frac{3  l_3 a}{2 v} + \eta s l_2 M \left(5+\frac{4}{v^2}\right)  \right],\\
    \gamma_{4}^{\text{KD}}=& \frac{\hat{q} \mu \sqrt{1-v^2}}{v} \left\{-\frac{2 l_4 a M}{v}\left(5+\frac{2}{v^2}\right)\right.\nn\\
    &+ 3 \eta  l_2 \left(-\frac{4a M}{v}+\eta \frac{\hat{q}\mu\sqrt{1-v^2}}{v}\right)\nn\\
    &\left.+\frac{9\eta s l_3}{10}\left[2a^2+M^2\left(18+\frac{45}{v^2}+\frac{5}{v^4}\right)\right]    \right\}.
\end{align}
\end{subequations}
The first three orders of the deflection \eqref{eq:kddef} coincide with the result from Ref. \cite{Li:2023svb}. For the latter usage, we can also define for the current case a truncated deflection angle $\Delta\phi_{\text{KD},\bar{n}}$ and the magnetic part $\Delta\phi^{\text{Mag}}_{\text{KD},\bar{n}}$ in a way completely analogous to Eqs. \eqref{eq:knnbar} and \eqref{eq:KN_zn_eta}.

\begin{figure}[htp!]
\centering
\includegraphics[width=0.45\textwidth]{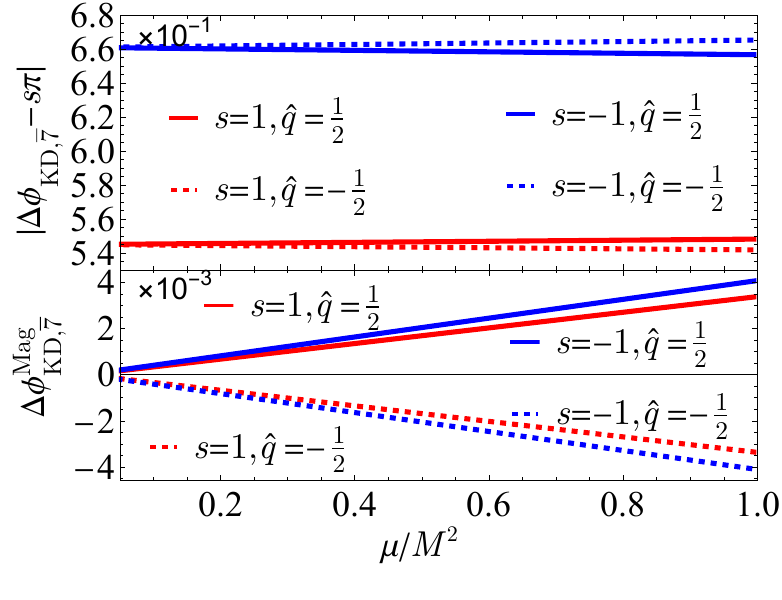}
\caption{$\Delta\phi_{\text{KD},\bar{7}}$ and the effect $\Delta\phi_{\text{KD},\bar{7}}^{\text{Mag}}$ of the magnetic field as functions of $\mu$ with other parameters set as $a=M/2,~v=1-10^{-2}, ~r_s=r_d=10^5M,~b=10M$, except for the running one. }
\label{fig:LiDeltaPhi}
\end{figure}

Different from the KN case in the previous subsection, the magnetic field here is characterized by the dipole moment $\mu$. This parameter is unique in that it participates only in the electromagnetic interaction but not in the gravitational one, in contrast to the spacetime charge $Q$ in the KN case. Before we study the effect of $\mu$ on the deflection, a simple analysis of the results \eqref{eqs:LDS_zn_Neu} and \eqref{eqs:LDS_zn_EM} suggests that the deflection is invariant under the sign change of $s,\,a,\,\mu$ and $\hat{q}$ as long as the signs of three products $sa$, $s\hat{q}\mu$ and $a\hat{q}\mu$ are unchanged. This degeneracy means that we can choose to study only the $(a>0,\,\mu>0)$ section of the parameter space without losing any generality. For example, a switching of the spin $a$ orientation can be compensated by the change of the rotation direction $s$ and the charge sign $\hat{q}$ in order to obtain the same deflection angle. 

In Fig. \ref{fig:LiDeltaPhi}, the deflection angle up to order seven and its magnetic part are plotted as functions of $\mu$. 
It is seen that in general for all combinations of $s$ and $\hat{q}$ signs, the effect of $\mu$ is still relatively weaker than the gravitational deflection itself, while the large separation between the red and blue curves is due to the gravitational effect of the spacetime spin $a$. 
For the parameter choice $(a>0,\,\mu>0)$ in the figure, the magnetic field has a component $B_\theta>0$, i.e., downward from the equatorial plane. Therefore, the Lorentz force is attractive if the charge $\hat{q}>0$ and the trajectory rotates counterclockwise, resulting in an increase to the deflection angle in this case, as seen from the red solid line. Dependence of the deflection angles of other combinations of $s$ and $\hat{q}$ signs on $\mu$ can be understood similarly. The same features as above can be more clearly seen from the lower panel of Fig. \ref{fig:LiDeltaPhi}, which shows separately the magnetic part of the deflection angle in this case. Note here that the blue solid line represents a repulsive Lorentz force, which effectively increases a negative total deflection angle (clockwise rotation) but actually reduces its absolute value, as seen from the top panel.

\subsection{Mass possessing magnetic dipole moment}

Besides the cases in the above subsections, a magnetostatic field can also be possessed by a mass without explicit involvement of a charge or current loop in its production. 
The line element of such a mass possessing a magnetic dipole moment is given by Ref. \cite{Gutsunaev:1987}, with metric functions on the equatorial plane as 
\begin{subequations}\label{eqs:PMDmetric}
\begin{align}
A\left(r\right)=&\left[1-\frac{2M(1+\alpha^{2})}{(1-3\alpha^{2}) r-4M\alpha^{2}}\right]\frac{K^{2}}{N^{2}}\nn\\
    =&1 -\frac{2M}{r}-\frac{16M^2\alpha^2\left(1+\alpha^2\right)}{\left(1-3\alpha^2\right)^2}\frac{1}{r^2}+\mathcal{O}\left[\frac{1}{r}\right]^3,\\
B\left(r\right)=&0,\\
C\left(r\right)=&\frac{1}{A(r)}\left[\left(r-M\right)^2-M^2\left(\frac{1+\alpha^2}{1-3\alpha^2}\right)^2\right]\nn\\
    =&r^2+\frac{8 \alpha^2 \left(3 \alpha^2+1\right) M^2}{\left(1-3 \alpha^2\right)^2}+\mathcal{O}\left[\frac{1}{r}\right]^1,\\
D\left(r\right)=&\frac{\mathrm{e}^{2\gamma}  \left(r-M\right)^2}{A(r)\left[\left(r-M\right)^2-M^2\left(\frac{1+\alpha^2}{1-3\alpha^2}\right)^2\right]}\nn\\
    =&1+\frac{2 M}{r}+\frac{4 \left(1+11 \alpha^4\right) M^2}{\left(1-3 \alpha^2\right)^2 r^2}+\mathcal{O}\left[\frac{1}{r}\right]^3,\\
\mathcal{A}_{\phi}\left(r\right)=&4 M^{2}\alpha^{3}(1-3\alpha^{2})^{-1}\frac{P}{K},  \label{eq:MDAphiRaw}
\end{align}
\end{subequations}
where the intermediate variables are
\begin{equation}
    \begin{aligned}
        K=&\left[\left(1-3\alpha^{2}\right)^{2}\left(r-M\right)^{2}-M^{2}\left(1+\alpha^{2}\right)\alpha^2\right]^{2}\\
        &+4M^{2}\alpha^{2}\left(1-3\alpha^{2}\right)^{2}\left(r-M\right)^{2},\\
        N=&\left\{\left[\left(1-3\alpha^{2}\right)\left(r-M\right)-M\alpha^{2}\right]^{2}+M^{2}\alpha^{2}\right\}^{2},\\
        P=&\left(1-3\alpha^{2}\right)^{3}\left(2r-M\right)\left(r-M\right)^{2}+M^{3}\left(1+\alpha^{2}\right)^{2}\alpha^{2},\\
        \mathrm{e}^{2\gamma}=&\frac{\left[\left(1-3\alpha^{2}\right)^{2}\left(r-M\right)^{2}-M^{2}\left(1+\alpha^{2}\right)^{2}\right]K^{4}}{\left(1-3\alpha^{2}\right)^{18}\left(r-M\right)^{18}}.
    \end{aligned}
\end{equation}
Here, the two important parameters are the spacetime mass $M$ and the magnetic parameter $\alpha$. The latter characterizes the magnitude of the magnetic field because it is related to the magnetic dipole moment $\mu$ through \cite{Gutsunaev:1987}
\be\label{eq:INeAlpha2Mu}
\mu = \frac{8 M^2 \alpha^3 }{\left(1-3 \alpha^2\right)^2 }.
\ee
Consequently, the asymptotic expansion of $\mathcal{A}_\phi$ in Eq. (\ref{eq:MDAphiRaw}) can be written using $\mu$ as
\begin{align}
    \mathcal{A}_{\phi}\left(r\right)=& \frac{\mu}{r}+\frac{3M\mu}{2r^2}+\frac{2M^2\mu\left(1-7\alpha^2+10\alpha^4\right)}{\left(1-3 \alpha^2\right)^2r^3}+\mathcal{O}\left[\frac{1}{r}\right]^4. \label{eq:mdaphi}
\end{align}
Note that when $\alpha=0$, the metric described by Eq. \eqref{eqs:PMDmetric} reduces to that of the Schwarzschild spacetime.

To find the deflection angle in this spacetime, inserting the expansion coefficients in Eqs. (\ref{eqs:PMDmetric}) and \eqref{eq:mdaphi} into Eqs. (\ref{eqs:zn}), the deflection angle in this case is found to be 
\begin{align}
\Delta \phi_{\text{MD}}=s\sum_{i=0}^{\infty} \frac{\beta_{i}^{\text{MD}}+\gamma_{i}^{\text{MD}}}{b^i}, \label{eq:PMDdef}
\end{align}
with the first few gravitational part coefficients
\begin{subequations}\label{eq:MDM_zn_Neu}
\begin{align}
    \beta_{0}^{\text{MD}}=&  l_0,\\
    \beta_{1}^{\text{MD}}=&   \left(1+\frac{1}{v^2}\right)M l_1,\\
    \beta_{2}^{\text{MD}}=& \left[\alpha^4 \left(\frac{59}{2}+\frac{70}{v^2}\right)+\alpha^2 \left(7-\frac{20}{v^2}\right)\right.\nn\\
    &\left.+\frac{3}{2}+\frac{6}{v^2}\right]\frac{M^2}{\left(1-3 \alpha^2\right)^2} l_2, \label{eq:MDM_zn_Neu2}
\end{align}
\end{subequations}
and the magnetic part coefficients
\begin{subequations}\label{eq:MDM_zn_EM}
    \begin{align}
    \gamma_{0}^{\text{MD}}=& 0,\\
    \gamma_{1}^{\text{MD}}=& 0,\\
    \gamma_{2}^{\text{MD}}=& \frac{ s \hat{q} \mu \sqrt{1-v^2}}{ v} l_1.\label{eq:MDM_zn_EM2}
\end{align}
\end{subequations}
Note that since this deflection angle is a product of power series of $M/b$ and  $1/\left(1-3 \alpha^2\right)$, the convergence of the series requires that $b > M/\left(1-3\alpha^2\right)$, which is always satisfied since $\alpha$ is usually much smaller than $1/\sqrt{3}$ due to the finiteness of the magnetic dipole moment in Eq. \eqref{eq:INeAlpha2Mu}.  

From coefficients \eqref{eq:MDM_zn_EM}, we observe no electrostatic contribution to the deflection, and the magnetic part is proportional to $\hat{q}$ at the leading order. Therefore, the signal charge $\hat{q}$ can function as the flag symbol $\eta$ used in Subsecs. \ref{subsec:KN} and \ref{subsec:kd}. After defining the truncated deflection angle $\Delta\phi_{\text{MD},\bar{n}}$ similarly to Eq. (\ref{eq:knnbar}), this enables us to define the magnetic contribution to the deflection as 
\begin{align}  \Delta\phi_{\text{MD},\bar{n}}^{\text{Mag}}=\Delta\phi_{\text{MD},\bar{n}}-\Delta\phi_{\text{MD},\bar{n}}|_{\hat{q}=0}.\label{eq:MDMmag}
\end{align}
We also observe that the dependence of the total deflection angle on $s$ and $\hat{q}$ is extremely simple in this case. The former provides only an overall sign of the deflection in Eq. \eqref{eq:PMDdef} and the latter participates in the magnetic contribution linearly as seen in Eqs. \eqref{eq:MDM_zn_EM}. 

\begin{figure}[htp!]
\centering
\includegraphics[width=.45\textwidth]{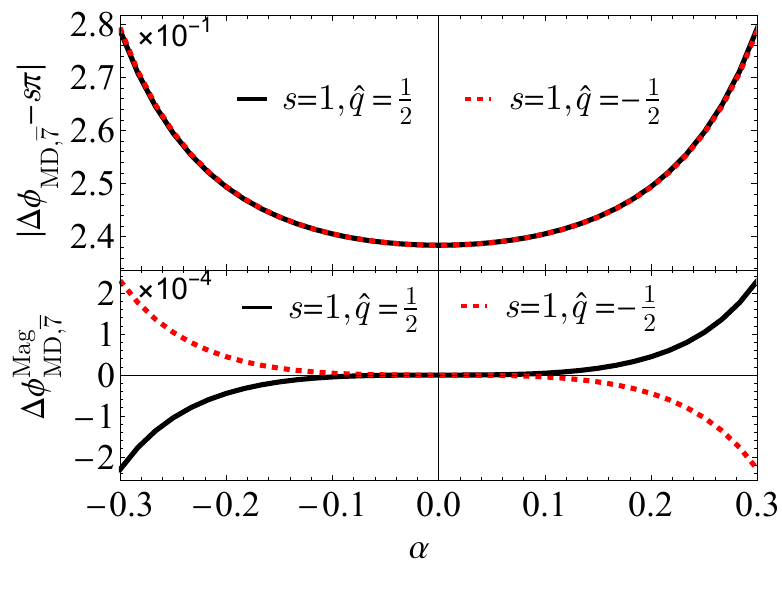}
\caption{$\Delta\phi_{\text{MD}, \bar{7}}$ and its magnetic part $\Delta\phi_{\text{MD}, \bar{7}}^{\text{Mag}}$ as functions of $\mu$. Other parameters used are $v=1-10^{-2}, ~r_s=r_d=10^5M,~b=20M$. We only plot the case $s=1$ because the case $s=-1$ is only a sign different. }
\label{fig:INeonMu}
\end{figure}

The only parameter affecting the deflection angle non-trivially in this case is $\alpha$. It is seen from Eqs. \eqref{eq:MDM_zn_Neu} and \eqref{eq:MDM_zn_EM} that it affects the deflection of charged particles through both the gravitational and magnetic channels, although its  square is directly proportional to the magnetic moment as in Eq. \eqref{eq:INeAlpha2Mu}. To see more clearly its effect, in Fig. \ref{fig:INeonMu} we plot, as functions of $\alpha$, the total deflection \eqref{eq:PMDdef} and the magnetic part $\Delta\phi_{\text{MD},\bar{n}}^{\text{Mag}}$. 
It is seen from the top panel of Fig. \ref{fig:INeonMu} that the total deflection has a relatively strong dependence on $\alpha$, mainly due to the gravitational part in Eq. \eqref{eq:MDM_zn_Neu2}. This total deflection is also approximately symmetric about the sign of $\alpha$, again a feature determined by the gravitational part of the interaction, Eqs. \eqref{eq:MDM_zn_Neu}. In contrast, from the lower panel we see that the magnetic part depends on $\alpha$ oddly, in agreement with Eq. \eqref{eq:MDM_zn_EM2}. One might expect that this oddity should be reflected in the top panel for the total deflection since the $\gamma_2^{\text{MD}}$ and $\beta_2^{\text{MD}}$ are both of order $M^2/b^2$. However, a careful inspection immediately shows that the $\gamma_2^{\text{MD}}$ contribution is much smaller due to the extra suppression from the reciprocal Lorentz factor $\sqrt{1-v^2}$. This also explains why it appears that the $\hat{q}=-1/2$ curve overlaps with the $\hat{q}=1/2$ one in the top panel, although in fact they do not. 

\section{Conclusion}\label{sec:conc}

A perturbative method is developed to assess the impact of the (electro)magnetic fields on the weak deflection angle of charged particles in the equatorial plane of general SAS spacetimes. The deflection angle can be solved to arbitrarily high order in the form of a series of $1/b$ as in Eqs. \eqref{eq:phiOnZ} and (\ref{eq:Deltaphi_on_b}), whose coefficients are determined by the asymptotic expansion coefficients of the metric functions. The finite distance effect of the signal source and observer is also incorporated in the factors $l_n$ of the series. 
Generally, it is found that the perturbative deflection angle at each order is always separated into two non-overlapping parts, the gravitational part and the electromagnetic part. The latter is suppressed by the reciprocal Lorentz factor but enhanced by the large specific charge, as seen from Eqs. \eqref{eq:phiOnZ} and \eqref{eq:Deltaphi_on_b}. Moreover, when there exist both electrostatic and magnetic interactions, the former can appear one order lower than the latter if the spacetime itself is charged. 

We then conducted a comprehensive analysis of deflection angles within three significant spacetimes as given in Eqs. (\ref{eq:KN_Neu})-(\ref{eq:KN_EM}), (\ref{eqs:LDS_zn_Neu})-(\ref{eqs:LDS_zn_EM}) and (\ref{eq:MDM_zn_Neu})-(\ref{eq:MDM_zn_EM}), and meticulously examined the influence of parameters associated with electromagnetic interactions on these angles.
Although in different scenarios the parameters affecting the magnetic contributions could be quite different, it is found that generally in all cases, a repulsive (or attractive) Lorentz force will decrease (or increase) the amount of deflection caused by gravity, in accord with simple intuition. This indeed is a reflection of the basic fact that what we did was a PN computation in the weak deflection limit, so that the Lorentz force does not overthrow the gravitational effect. If the magnetic field is allowed to be stronger (e.g., a uniform magnetic field) or the impact parameter could approach a much smaller value, it is expected that these conclusions could change.

A potential direction of extension of the current work involves the treatment of a uniform magnetic field for its deflection of charged signals. The methodology presented in this study is currently inapplicable to a uniform magnetic field, as its expansion of $\mathcal{A}_\phi$ at infinity begins from $r^2$. Nevertheless, a very weak uniform magnetic field that extends throughout space holds both theoretical importance and computational viability. The second extension of the current work is to apply the current method to more interesting spacetimes or gravitational theories with magnetic fields, revealing the effect of spacetime/magnetic field parameters in those cases on the deflections, hoping to enhance the chance to confirm or falsify such spacetimes or gravitational theories. We are currently working along some of these directions. 

\acknowledgements

The authors appreciate the discussion with Qian Li, Yuhan Zhou and Jinhong He. This work is partially supported by the Wuhan University Research Development Fund.

\appendix

\section{$l_n$ and their limits}\label{sec:ln}

As Eq. (\ref{eq:ln}), we can calculate $\displaystyle l_n(\delta_s,\delta_d)$ as in Eq. (2.16) of Ref.  \cite{Huang:2020trl}. The results are
\begin{subequations}
    \begin{align}
     l_0=&\sum_{i=s,d}\left(\frac{\pi}{2}-\delta_i\right),\\
        l_1=&\sum_{i=s,d}\left(\cos \delta_i\right),\\
        l_2=&\sum_{i=s,d}\left(\frac{\pi}{4}-\frac{1}{2}\delta_i +\frac{1}{4}\sin 2\delta_i\right),\\
        l_3=&\sum_{i=s,d} \left(\frac{2 }{3}\cos \delta_i + \frac{1}{3}\cos \delta_i \sin^2 \delta_i\right),\\
        l_n=&\sum_{i=s,d}\left[\frac{\cos (\delta_i) \sin ^{n-1}(\delta_i)}{n}\right] +\frac{n-1}{n} l_{n-2}\left(\delta_s,\delta_d\right) \nn\\
        &~~~~~~~ \text{for} \quad n \geq 2.
    \end{align}
\end{subequations}
Here $\delta_s$ and $\delta_d$ are related to the finite value of $r_s$ and $r_d$ through definition \eqref{eq:apperentangle}. With the asymptotic expansions of the metrics and the electromagnetic four-potential in Eqs. \eqref{eq:metricCoe}, $\delta_{s,d}$ can also be expanded, to find 
\begin{align}
    \delta_i=&\frac{b}{r_i}+ \left[ -\frac{s}{v}\left(\frac{b_1}{2}-\hat{q}\mathfrak{q}_{31}\sqrt{1-v^2}\right) \right.\nn\\
    &\left.+b\left(-\frac{c_1}{2}+\frac{a_1-2\hat{q} \mathfrak{q}_{01}\sqrt{1-v^2}}{2v^2}\right) \right]  \frac{1}{r_i^2}+\mathcal{O}\left(\frac{1}{r_i^3}\right).
\end{align}
Clearly, in the infinite source and detector distance limit, $\delta_{s,d}\to 0$ and therefore $l_n$ reduce to
    \begin{align}
     l_0=&\pi,\quad
        l_1=2,\quad
    l_2=\frac{\pi}{2},\quad
        l_3=\frac{4}{3},\nn\\
        l_n=&\frac{n-1}{n} l_{n-2}~~ \text{for} ~~ n \geq 2.
    \end{align}

\end{document}